%Paper: cond-mat/9406086
%From: John Neergaard <neergard@u.washington.edu>
%Date: Tue, 21 Jun 1994 16:57:19 -0700 (PDT)

\magnification= 1200

\def\Ham{{\cal H }}
\def\Trans{{\cal T }}

\def\half{{ 1\over 2 }}

\def\3half{{3\over 2 }}

\hsize= 6.0 true in
\vsize= 8.5 true in
\baselineskip 18 pt plus 0.2 pt minus 0.2 pt

\centerline{\bf CROSSOVER SCALING FUNCTIONS IN}
\centerline{\bf ONE DIMENSIONAL DYNAMIC GROWTH MODELS }
\bigskip
\centerline{\bf John Neergaard and Marcel den Nijs}
\bigskip
\centerline{\it Department of Physics}
\centerline{\it University of Washington}
\centerline{\it Seattle, WA 98195}
\vskip 50 pt
\noindent
\baselineskip 15 pt plus 0.2 pt minus 0.2 pt

The crossover from  Edwards-Wilkinson ($s=0$) to
KPZ ($s>0$) type growth is studied for the BCSOS model.
We calculate the exact numerical values for the $k=0$ and  $2\pi/N$ massgap
for $N\leq 18$  using the  master equation. We predict the structure of the
crossover scaling function and confirm numerically that
$m_0\simeq 4 (\pi/N)^2 [1+3u^2(s) N/(2\pi^2)]^{0.5}$ and
$m_1\simeq 2 (\pi/N)^2 [1+ u^2(s) N/\pi^2]^{0.5}$,
with $u(1)=1.03596967$.
KPZ type growth is equivalent to  a phase transition in meso-scopic
metallic rings
where attractive interactions destroy the persistent current;
and to endpoints of facet-ridges in equilibrium crystal shapes.
\vfill
\eject
\baselineskip 18 pt plus 0.2 pt minus 0.2 pt

A large amount of theoretical effort has been devoted in recent years
to establish and classify the scaling properties of dynamic processes
such as those at growing interfaces.
Several dynamic universality classes have emerged.
One of them is the so-called KPZ universality class,
named after the non-linear Langevin equation studied
by Kardar, Parisi, and Zhang [1].
Older examples are Edwards-Wilkinson (EW) growth [2]
and directed percolation [3].
They are distinguished by the value of
the dynamic critical exponent $z$;
characteristic lengths and times  scale as $t \sim l^z$.
Most of the evidence is numerical in nature,
in particular from Monte Carlo (MC) studies [2], supplemented by a few exactly
soluble models [4]  and field theoretical renormalization
studies [1,2].

Our understanding  of dynamic scale invariance is still nowhere near that
of equilibrium critical phenomena.
In particular the comparison with two dimensional (2D) critical phenomena
is relevant.
1D dynamic processes and  2D equilibrium statistical mechanics are very
much alike
in the master equation and transfer matrix formulation.
Time becomes the second spatial coordinate.
Most  2D equilibrium critical phenomena obey conformal invariance [5].
They all belong  to the same conformal ``dynamic" universality class ($z=1$;
rotational invariance).
The ground state properties of 1D quantum systems are related as well.
Luttinger liquids (relativistic fermions) are also part of the conformal
$z=1$ dynamic universality class [6].
Non-relativistic fermions have a dynamic exponent $z=2$.
Anisotropic scaling, with $z\neq1$, occurs in 2D critical phenomena at  e.g.
Pokrovksy-Talapov (PT) transitions ($z=2$) and the elusive  Lifshitz points
[6].
One prospect is that concepts similar to conformal invariance
might apply to dynamic universality classes in general.

In this letter we present a detailed numerical study of the finite size scaling
(FSS) crossover function  from EW ($z=2$) to KPZ ($z=1.5$) type growth in the
body-centered solid-on-solid (BCSOS) growth model.
It is important to know the details of the crossover scaling
between dynamic universality classes.
The following simple example illustrates this.

Consider the 1D free fermion model.
The fermions are non-relativistic
when the chemical potential coincides with the bottom of the energy band
$E(k)\simeq k^2$.
The momenta are spaced as $\Delta k = 2\pi/N$.
So the massgap scales as $m\simeq (2\pi/N)^2$ and
$z_0=2$ because  $l_t\sim m^{-1}$.
The fermions are relativistic for $k_F>0$.
The massgap scales as $m\simeq 4\pi k_F/N$, and $z_1=1$.
The crossover scaling function
$m=(k_F+2\pi/N)^2-k_F^2\simeq 4\pi k_F/N +(2\pi/N)^2$
has two important properties.
The crossover exponent $y_c$ determines the  scaling of
$dm/dk_F\sim N^{-z_0+y_c}$ at $k_F=0$.
$y_c$ is equal to the change  in the dynamic exponent, $y_c =z_0-z_1 = 1$.
The FSS amplitude of the massgap is not a constant in the metal phase,
but a universal number multiplied with the rapidity $2k_F$.
The metal phase belongs to the conformal  $z=1$ dynamic universality class.
The rapidity represents the lattice anisotropy in 2D critical
phenomena language.
The crossover operator is the stress tensor which lies at the heart
of conformal field theory.
We expect that
crossover scaling functions between dynamic universality classes
have the same generic assymptotic form
$$
m \simeq A/N^{z_0} + B u(s)/N^{z_1},
\eqno(1)
$$
with $A$ and $B$ universal constants and $u(s)$ the scaling field
associated with the crossover operator.

Surface roughness is characterized by the scaling of the
height-height correlation function,
$
G(r,t) = < [h(r_o+r)-h(r_0)]^2 > \simeq  b^{2\alpha} G(b^{-1} r, b^z t^{-1}).
$
In 1D the stationary state has typically a finite correlation length
beyond which
the steps are disordered, and the probability to go up or down
along the interface becomes random.
At large length scales $G(r)$ behaves like
the root mean-square displacement of a random walk, $G(r) \sim r$.
So typically $\alpha=0.5$.
Stationary states with interesting correlations like those in 2D
driven diffusive systems [7] have not been found (yet) in 1D growth dynamics.
$z$ characterizes the rate at which the stationary state is approached.
A disturbance with characteristic size $\xi$
spreads-out as $\xi \sim t^z$.

In the  BCSOS model nearest neighbour columns
must differ in height by one unit, $dh=\pm1$.
This describes a body-centered type stacking of molecules,
or a zig-zag type stacking of bricks.
We represent the surface configuration in terms of the steps, $S^z_n=\pm1$.
The dynamic rule is as follows.
Select one of the columns at random;
if $S^z_n=-1$ and $S^z_{n+1}=1$ (a local valley)
a brick is adsorbed  with probability $p$;
if $S^z_n=1$ and $S^z_{n+1}=-1$ (a local hill top)
evaporation occurs  with probability $q$.
$P(\{S^z_n\})$ is the probability of micro state $\{S^z_n\}$.
$|\Psi\rangle _{\tau} =$ $\sum P(\{S^z_n\}) |\{S^z_n\} \rangle$
is the state vector.
The time evolution operator $\Trans$ of the discrete-time master equation
$|\Psi\rangle _{\tau+\Delta}= \Trans |\Psi\rangle _{\tau}$
has the following structure
$$
\eqalignno{
\Trans = 1 -
{1 \over 4} \epsilon  N^{-1}  \sum _n [ 1 -\lambda S^z_n S^z_{n+1}
- \half   &(S^+_n S^-_{n+1} + S^-_n S^+_{n+1}) \cr
- \half s &(S^+_n S^-_{n+1} - S^-_n S^+_{n+1}) + \mu S^z_n] &
(2)\cr}
$$
with
$\Delta=1/N$,
$s = (p-q)/(p+q)$,
and periodic boundary conditions, $S^z_{n+N}=S^z_n$.
Without loss of generality time can be rescaled to $\epsilon= p+q=1$.
The parameters $\lambda =1$ and  $\mu =0$ are introduced for later convenience.
Stochastic processes  preserve probability.
So the state where all microstates have the same probability,
$|D\rangle =\sum  |\{S^z_n\}\rangle$,
is the left eigenvector for the largest eigenvalue, $\lambda_0 =1$.

The BCSOS model has more symmetry than  generic KPZ growth models:
$\Trans(p,q) = \Trans(q,p)^\dagger$.
The left eigenvectors $\langle L|$  of $\Trans(p,q)$
are identical to the  right eigenvectors  of $\Trans(q,p)$.
$\Trans (q,p)$ and $\Trans(p,q)$ are equivalent by  particle-hole symmetry.
Therefore $ | R\rangle = ~[\prod S^x_n]~|L\rangle $,
and $|D\rangle$ is both the left and the right eigenvector for $\lambda_0$.
The stationary state is completely disordered;
$\alpha=1/2$  for all $s$.
The BCSOS model is used  widely to study  KPZ type growth [2].
This special symmetry casts some doubts on its  role as
a generic KPZ growth model.
This  becomes more apparent in the crossover scaling behaviour.

It is important to point out how this model fits into the general picture
sketched above.
$\Trans$ has the generic form, $\Trans = \exp (- \Ham )$.
$\Ham$ is non- Hermitian, except at $s=0$ where the model
reduces to the spin-1/2 XXZ chain [6].
The EW point  at $s=0$ maps exactly onto the ferromagnetic
Heisenberg point, $\lambda=1$.
In the KPZ equation EW growth corresponds to the point where
the non-linear term vanishes and the dynamics reduces
to a stochastic diffusion equation.
Indeed, the exact solution of the  XXZ chain at $\lambda=1$ is diffusive.
The equation of motion for the spin-spin correlation
closes and is a deterministic diffusion equation [8].
We checked that also the equations of motion
for the $n$-point spin correlation functions close within themselves
and are diffusion type equations.

The XXZ model is equivalent to a 1D spinless interacting fermion problem [6].
Growth,  $s$, introduces a preferred hopping direction.
It creates a persistent current around the ring,
similar to those in metallic rings in perpendicular magnetic fields.
Recent research is focused on coherence effects in the presence of
elastic scattering from random impurities [9].
Eq.(2) does not include impurities.
It describes a  phase transition due to interactions.
Weak attraction between the fermions, small $\lambda$,
only renormalizes the scaling indices
due to the marginal operator in the chiral Luttinger liquid.
At  $\lambda>1$ the fermions coalesce into a macroscopic bound state
(the ferroelectric ground state of the XXZ chain).
This destroys the persistent current, because in
eq.(2)  only single fermions can hop.
(In metallic rings the macroscopic droplets of fermions might still be
able to move).
The transition takes place at the KPZ line, at $\lambda=1$.
At this transition the fermion excitations do not behave metallic, $z=1$,
but are described by the unconventional dynamic exponent $z=1.5$.
It will be interesting to investigate how elastic impurities modify this.

The spin-$1/2$ XXZ chain is related to
the  6-vertex model for 2D equilibrium (anti-) ferroelectrics [10].
The equivalence becomes exact
when in the dynamic rule the collumns are chosen sequentually instead
of randomly, first  all even and then all odd collumns [8].
The modified rule reduces to the old one in the time continuum limit,
$p\to 0$ and  $q\to 0$ with $s$ constant.
$\mu$ and $s$ are the two components of the electric field.

The 6-vertex model is equivalent to
the BCSOS model for equilibrium crystal surfaces.
The anti-ferroelectric  side of its phase diagram describes
surface roughening transitions [10].
The ferroelectric side is less well known.
It describes the crystal shape close to a facet ridge [11,12].
$\lambda$ plays the role of inverse temperature.
At $\lambda>1$ two facets meet at a facet ridge; see Fig.~1.
The growth parameter $s$ tilts the surface in the time-like direction,
the direction along the facet ridge.
The chemical potential $\mu$ tilts the surface in the spatial direction,
the direction $\perp$ to the facet ridge.
The interactions in this direction are anti-ferrmagnetic and favour facetting.
The KPZ points are the endpoints  of the  facet ridge
where the two facets become separated by two PT transition lines
with rough rounded surface in-between; see Fig.~1.
The KPZ line represents the change in location
with temperature of these facet ridge endpoints.
The facet ridge shortens with  temperature
until it vanishes at the EW critical point.
At PT transition  the surface rounds  smoothly.
At facet ridge endpoints it changes discontinuously, because
at KPZ points the surface has a finite slope in the time-like direction
equal to the growth rate $v_g=s/4$.
The dynamic exponent $z=1.5$ implies unusual anisotropic scaling behaviour.
There is one caveat to this identification of KPZ growth with
facet ridge endpoints.
More general growth models such as the RSOS model lack
the self-adjointness property of eq.(2).
Moreover, at the facet ridge endpoints in the BCSOS model the
transfer matrix happens to be stochastic;
we see no reason for this to be true in general.
It will be interesting to investigate how these aspects
affect the scaling properties.

The eigenvalues of the transfer matrix can be classified according
to wave numbers $k$, reflecting translational invariance.
The largest eigenvalue, $\lambda_0=1$, is located in the $k=0$ sector.
Define two massgaps: $m_0= \lambda_0(0)-\lambda_1(0)$
and $m_1= \lambda_0(0)-\lambda_0(1)$.
$\lambda_1(0)$ is the next largest eigenvalue
in the $k=0$ sector, and $\lambda_0(1)$
the largest eigenvalue in the $k =2\pi/N$ sector.
We calculated  the exact numerical values of
$m_0$ and $m_1$ for even system sizes  $N \leq 18$.
The dynamic exponent follows from the FSS of the massgaps,
$m\sim N^{-z}$.
The spatial exponent follows independently
from the interface width, $W\sim N^\alpha $,
in the stationary state (the right eigenvector of $\lambda_0(0)$).

FSS approximants for  $z$ are shown in Fig.~2 at various values of $s$.
$z(N,N+2) = \ln(m_1(N)/m_1(N+2)) /\ln((N+2)/N)$.
These numbers are accurate to better than 12 decimal places,
unlike results from e.g. MC simulations.
Therefore it is possible to take into  account the leading corrections
to scaling.
We do this by constructing a cascade of $1/N$ extrapolations
for the $z(N,N+2)$  at successive values of $N$.
At $s=1$ the convergence is very stable, and leads to $z=1.50\pm0.05$.
This is the same level of accuracy as obtained by  MC simulations,
but with substantially less effort and much smaller system sizes.

The entire line  $0<s\leq1$ belongs to the KPZ universality class.
The dynamic exponent must be equal to $z=1.5$ for all $s\neq0$,
but the convergence is only good for $s\simeq 1$.
For $s\to 0$, the $z(N,N+2)$  move towards
the  EW value $z=2$ at small $N$ before  they bend-down towards
the KPZ value $z=1.5$.
This reflects the crossover scaling behaviour from the nearby EW point.

The  crossover scaling function eq.(1) does not apply to the BCSOS model,
due to the special symmetry discussed above, which implies that
the massgaps are even functions of $s$.
Their first derivative vanishes.
Eq.(1) is likely to apply to the more generic growth models,
but this needs still to be tested.
We propose the modified form for the BCSOS model:
$$
m = {A\over N^2}[1+ B {u^2(s)~N}]^{0.5}.
\eqno(3)
$$
Also in this expression $y_s$  is equal to the change in the dynamic exponent,
and the amplitude at $s\neq 0$ is equal to a universal number multiplied
with $u(s)$.

The crossover scaling exponent follows
from the second derivatives of the massgaps with respect to $s$,
$m^{\prime\prime} \sim  N^{-z+2y_s}$.
We find $y_s = 0.50 \pm 0.05$;
consistent with $y_s=0.5$ from power counting at the EW point
in the KPZ equation.

At $s=0$ the $k=2\pi/N$ mass gap is exactly equal to
$m_1(0)=2 \sin^2(\pi/N)$ for all $N$.
Eq.(2) reduces to the ferromagnetic Heisenberg model.
Spin rotations leave $\Trans$ invariant, but  mix-up
the sectors of different magnetization (slopes of the surface).
Since  $\lambda_0(k)$   is non-degenerate in each sector
it follows that $m_1$  is independent of the tilt of the surface.
The equation for $m_1(0)$ is trivially true in
one particle sector with surface tilt $1-1/N$.

The $k=0$ massgap at $s=0$ is more complex.
We find numerically:
$
m_0(0)= 4 \sin^2(\pi/N) [1+ {2\over\pi} \sin(\pi/N) +(0.306 \pm 0.003)
\sin^2(\pi/N)+...]
$.
The exact result for $m_1(0)$ suggests that $\sin(\pi/N)$ is the natural
FSS parameter.
Only  FSS corrections with integer powers of $\sin(\pi/N)$ are to be
expected, because
all n-point correlation functions at $s=0$ obey diffusion type equations.
We find no numerical evidence otherwise.
We expect amplitudes to be simple combinations of $\pi$ and integers.
The leading correction to scaling amplitude of $m_0(0)$,  $0.635\pm 0.005$,
is close enough to $2/\pi$ to guess this to be the exact value.
Similarly, the second derivatives of the massgaps at $s=0$ scale
numerically as:
$
m_1^{\prime\prime}(0) =
{2\over \pi} \sin(\pi/N) -\sin^2(\pi/N)+(0.25\pm0.02)\sin^3(\pi/N) +...]
$
and
$
m_0^{\prime\prime}(0) =
{6\over \pi} \sin(\pi/N) - (1.06\pm0.01) \sin^2(\pi/N)+...]
$.
These results yield the amplitudes in eq.(3):
$$
\eqalignno{
m_1(s,N) &= 2 \sin^2(\pi/N) [1+{1\over \pi} u(s)^2/\sin(\pi/N)]^{0.5}\cr
m_0(s,N) &= 4 \sin^2(\pi/N) [1+{3\over2 \pi} u(s)^2/\sin(\pi/N)]^{0.5}&
(4)\cr
}
$$

At $s=1$ the leading amplitude of the $k=2\pi/N$ massgap is known exactly
from the Bethe ansatz solution,
$m_1 \simeq  1.1689666 (\pi/N)^\3half$ [4].
We use this in our scaling analysis at $s=1$:
$
m_1(1)= 1.1689666 \sin^\3half(\pi/N) [1+ (0.531\pm 0.002) \sin(\pi/N)+...]
$.
This gives information about the scaling field $u(s)$.
The renormalization is small.
The  amplitude predicted by eq.(4) assuming $u=s$ is only $7\%$ too small.
Our knowledge of the non-linear scaling field can be summarized as
$u(s)= s(1+ 0.0359697 s^2 f(s^2))$,
with $f(s^2)$ an unknown function, except for $f(1)=1$.

The amplitude of the $k=0$ massgap at s=1 is not
known from the Bethe Ansatz solution [4].
The two massgaps must depend on  $u(s)$ in the same way.
Therefore eq.(4) predicts the exact value,
$m_0(1) \simeq 2.8633717\sin^\3half(\pi/N)$.
This agrees well with a free numerical estimate,
$m_0(1) \simeq (2.87\pm 0.01) \sin^\3half(\pi/N)$.
Moreover, the leading corrections to scaling amplitude is very stable when
we assume this value to be exact:
$
m_0(1) \simeq 2.8633717 \sin^\3half(\pi/N) [ 1+(0.940\pm 0.005)
\sin(\pi/N) +...]
$.

It is tempting to visualize the low lying excitations in terms of
quasi-particles
with a dispersion relation $E(k)\sim k^z$.
The amplitude of the massgap $m_1\simeq A N^{-z}$ yields the slope
of such a dispersion relation.
In this picture,  $m_0$ must have an amplitude twice a big as $m_1$,
because the lowest $k=0$ excitations
involves the creation of two quasi particles of opposite momentum.
This is true at the EW point, but not for KPZ type growth where
the amplitudes differ by more than a factor 2.
So the large length scale behaviour of KPZ growth
can not be described by a free field theory this simple in structure.

Fig.~3 shows the convergence of the ratio between the numerical values
of $m_1(s)$ and those predicted by eq.(4),  as function of $s$.
The leading corrections to scaling are incorporated.
The representation of corrections to scaling is not unique
because their origins cannot be distinguished numerically.
We choose to represent them as:
$$
m_1 =
2 \sin^2({\pi\over N}) [ 1 +
{1\over \pi} {u^2(s) \over \sin(\pi/N)}
-b_1(s)s^2 +...]^{0.5}
$$
with $b_1(s) \simeq \half + 0.137 s^2$
from the corrections to scaling in $m^{\prime\prime}_1(0)$ and $m_1(1)$.
The ratio is a sensitive test
because both  quantities vanish in the thermodynamic limit.
The $k=2\pi/N$ massgap ratio converges to 1 within $0.1\%$ for all $s$.
Similarly, the  $k=0$ massgap ratio converges to within $0.5\%$.
The quality  of the convergence  varies slightly with $s$.
This reflects the uncertainty in the scaling field $u(s)$.
The shape of the curves is sensitive to the precise choice for $f(s^2)$.
$f(s^2)=1$ can be ruled out,  because our
numerical estimates of the amplitudes at $s=0.6$ is slightly
but significantly smaller
than predicted by $f(s^2)=1$.
We investigated  the simple form $f(s^2)=(1-a)s^2-a$ for various values of $a$.
$f(s^2)=s^2$, shown in Fig.~3,  gives the smoothest curves,
but the differences are marginal.
The convergence of both ratios stays within the limits quoted above
for any of these choices.

We conclude that the asymptotic crossover scaling function eq.(3)
is correct. The agreement with the exact numerical results
is within $7\%$ over the entire range $0\leq s\leq1$
when $s$ is assumed to be a pure scaling field.
The agreement improves to within $0.5\%$
when the renormalization of the growth parameter is incorporated.

This research is supported by NSF grant DMR-9205125.

\vfill
\eject
\centerline{\bf Figure Captions}
\item{  Fig.1:}
Schematic temperature evolution of a facet ridge in an equilibrium
crystal.
KPZ growth  coincides  with the
facet edge endpoint (a) and  EW growth
with the point where the facet ridge vanishes (b).
\item{  Fig.2:}
Finite size scaling approximants of the dynamic exponent $z$ at various
values of $s$
from the $k=2\pi/N$ massgap.
\smallskip
\item{  Fig.3:}
The ratio between the numerical values of the $k=2\pi/N$ massgap and
those of the proposed crossover scaling function eq.(4)
for system sizes $N=10-18$.
\vfill
\eject

\centerline{\bf references}
\item{ 1. } M.~Kardar, G.~Parisi, and Y.C.~Zhang, Phys.Rev.Lett. {\bf 56},
889 (1986).
\item{ 2. }
F.~Family, Physics A {\bf 168}, 561 (1991);
J.~Krug an H.~Spohn in
{\it Solids Far from Equilibrium; Growth, Morphology, and Defects},
edt. C. Godr\`eche (Cambridge University Press, Cambridge 1991);
J.~Villain,  J.Phys.Fr.I {\bf 1}, 19 (1991).
\item{ 3. }
W.~Kinzel in
{\it Percolation  Structures and Processes},
eds. G.~Deutscher, R.~Zallen, and J.~Adler
(Adam Hilger, Bristol, 1983).
\item{ 4. }
L-H.~Gwa and H.~Spohn, Phys.Rev.A {\bf 46}, 844 (1992).
\item{ 5. } J.~Cardy in
{\it Phase Transitions and Critical Phenomena};
eds. C.~Domb and J.~Lebowitz
(Academic, London,1987, Vol.11).
\item{ 6. } M.~den Nijs in
{\it Phase Transitions and Critical Phenomena};
eds. C.~Domb and J.~Lebowitz
(Academic, London,1988, Vol.12).
\item{ 7. } B.~Schmittmann and R.K.P.~Zia in
{\it Phase Transitions and Critical Phenomena},
eds C.~Domb and J.L.~Lebowitz
(Academic, London, Vol.17, in press).
\item{ 8. } E.~Domany, W.~Kinzel, and B.~Nienhuis, J.Phys.A {\bf 23},
L755 (1990).
\item{ 9. } F.~von Oppen and E.K.~Riedel, Phys.Rev.~B {\bf 48}, 9170 (1993).
\item{10. } I.M.~Nolden, J.Stat. Phys. {\bf67} 155, (1992).
\item{11. } J.~Neergaard and M.~den Nijs,  Bull.Am.Phys.Soc. {\bf38},
798 (1993).
\item{12. } J.D.~Shore and D.J.~Bukman, Phys.Rev.Lett. {\bf 72}, 604 (1994).

\vfill
\eject
\end